\documentclass[conference]{IEEEtran}
\usepackage{amsmath,amsfonts}
\IEEEoverridecommandlockouts
\usepackage{array}
\usepackage[caption=false,font=normalsize,labelfont=sf,textfont=sf]{subfig}
\usepackage{textcomp}
\usepackage{booktabs}
\usepackage{stfloats}
\usepackage{url}
\usepackage{verbatim}
\usepackage{graphicx}
\usepackage{hyperref}
\usepackage[ruled,linesnumbered]{algorithm2e}
\usepackage{cite}
\begin{document}

\title{EFILN: The Electric Field Inversion-Localization Network for High-Precision Underwater Positioning}

\author{
\IEEEauthorblockN{
Yimian Ding\thanks{$^+$ These authors contribute equally to this work. }\IEEEauthorrefmark{1}$^{,+}$,
Jingzehua Xu\IEEEauthorrefmark{1}$^{,+}$,
Guanwen Xie\IEEEauthorrefmark{1}, Haoyu Wang\IEEEauthorrefmark{2}, Weiyi Liu\IEEEauthorrefmark{1},
Yi Li\IEEEauthorrefmark{1}
}
\IEEEauthorblockA{\IEEEauthorrefmark{1}Tsinghua Shenzhen International Graduate School, Tsinghua University, Shenzhen, 518055, China}
\IEEEauthorblockA{\IEEEauthorrefmark{2}Ocean College, Zhejiang University, Hangzhou, 316021, China}
Email: liyi@sz.tsinghua.edu.cn
\thanks{
$^{1}$$\;$Codes are available at https://github.com/Xiboxtg/EFILN.}
}

\markboth{IEEE Wireless Communications and Networking Conference}%
{Shell \MakeLowercase{\textit{et al.}}: A Sample Article Using IEEEtran.cls for IEEE Journals}


\maketitle

\begin{abstract}
Accurate underwater target localization is essential for underwater exploration. To improve accuracy and efficiency in complex underwater environments, we propose the Electric Field Inversion-Localization Network (EFILN), a deep feedforward neural network that reconstructs position coordinates from underwater electric field signals. By assessing whether the neural network’s input-output values satisfy the Coulomb law, the error between the network's inversion solution and the equation's exact solution can be determined. The Adam optimizer was employed first, followed by the L-BFGS optimizer, to progressively improve the output precision of EFILN. A series of noise experiments demonstrated the robustness and practical utility of the proposed method, while small sample data experiments validated its strong small-sample learning (SSL) capabilities. To accelerate relevant research, we have made the codes available as open-source$^{1}$.
\end{abstract}


\section{Introduction}\label{se:1}

\IEEEPARstart{I}{n} recent years, precise and efficient underwater target localization has become a significant challenge\cite{khalil2020toward}. Most research on electric field inversion and localization has focused on traditional methods\cite{pan2021iout, su2022algorithm}. Lee \textit{et al}. \cite{lee2018real} proposed a real-time localization approach using a precomputed direct current (DC) electric field template. Ji \textit{et al}. \cite{7382734} simulated a moving dipole’s electromagnetic field and introduced a Kalman filter-based localization algorithm. Yu \textit{et al}. \cite{Yu2019SHIPTT} used an unscented particle filter to track ship-generated electric fields. However, these methods suffer from lower accuracy and limited applicability.

With advances in computer performance and forward modeling, intelligent nonlinear localization methods have emerged. Unlike traditional approaches, these methods address nonlinear inversion without relying on initial models or Jacobian matrices \cite{liu2021two}. Algorithms like simulated annealing \cite{shi1998one, sharma2012vfsares}, particle swarm optimization \cite{shaw2007particle, shi2009damped}, and genetic algorithms \cite{schwarzbach2005two} have been applied to electric field inversion. However, they often suffer from low accuracy and slow convergence with large datasets, limiting their effectiveness.

Recently, deep learning (DL) algorithms have attracted attention for their potential in electric signal inversion\cite{10650551,inproceedings}. Puzyrev \textit{et al}.\cite{puzyrev2019deep} used convolutional neural networks (CNNs) for two-dimensional (2D) inversion of controlled-source electromagnetic data\cite{di2020new}. Liu \textit{et al}.\cite{liu2020deep} applied CNNs for 2D inversion of electrical resistivity data, mapping resistivity to geoelectric models. These studies highlight DL's potential in signal inversion, but few have explored its use in electric signal localization.

Building on the above analysis, this paper proposes a novel deep learning neural network, termed the Electric Field Inversion-Localization Network (EFILN), specifically designed for high-precision and efficient electric field inversion localization. The proposed framework takes normalized three-directional underwater electric field components as inputs and outputs the corresponding position coordinates. By incorporating the physical principles of the electric field, governed by the Coulomb law, into the loss calculation, the output of EFILN gradually converges to the true values, achieving accurate electric field inversion localization. The primary contributions of this paper are summarized as follows
\begin{itemize}
\item{To the best of our knowledge, this is the first study to employ the deep learning model for target localization by inverting electric field signals based on the Coulomb law. This technique elucidates the relationship between electric field signals and target position coordinates, facilitating precise localization.}
\item{We employ a joint optimization approach using Adam and L-BFGS to update the neural network parameters. Adam rapidly reduces the loss during the initial training phase, allowing the model to quickly reach a promising parameter region. In the later stages, when the loss function approaches convergence, L-BFGS fine-tunes parameters with more accurate second-order information, further enhancing the model's performance. This approach improves both training efficiency and effectiveness.}
\item{Extensive simulation experiments demonstrate that EFILN can achieve high-precision electric field inversion localization in environments with varying noise intensities, exhibiting strong robustness and generalizability. Additionally, EFILN performs well even under small sample learning (SSL) conditions.}
\end{itemize}

% \begin{table}[!t]
% \centering
% \caption{Main Symbols and Explanations.\label{tab1}}
% \label{tab:1}
% \setlength{\tabcolsep}{2.0mm}{
% \begin{tabular}{c|c}
% \hline
% {\bf Symbols} & {\bf Definition}\\
% \hline\hline 
% $\phi$, $\phi^{\prime}$ & Source dipole moments\\
% \hline
% $A_j(x,y,z)$, $F_j(x,y,z)$ & Shelkunoff vector potentials\\ 
% \hline
% $E_{xj}$, $E_{yj}$, $E_{zj}$ & Electric field\\
% \hline
% $H_{xj}$, $H_{yj}$, $H_{zj}$ & Magnetic field\\
% \hline
% $G$ & Gaussian random noise\\
% \hline
% $\mathcal{F}_a(\cdot)$ & Activation function\\
% \hline
% $\mathcal{G}$ & The gradient\\
% \hline
% $\mathcal{L}(\cdot)$ & The loss funcion of the CNN model\\
% \hline
% $\mathcal{B}_1$, $\mathcal{B}_2$ & The exponential decay rates for moment estimates \\
% \hline
% $\mathcal{A}$, $\varsigma$ & The first and second moment variables\\
% \hline
% $\boldsymbol{\pi}$ & Policies of all the AUVs\\
% \hline
% $\boldsymbol{x}$ & Observations of all the AUVs\\
% \hline
% $H_0$ & The entropy\\
% \hline
% $Q_{\Theta_i}^{\pi_i}$ & The action value function\\
% \hline
% $\Theta_{1_i}$, $\Theta_{2_i}$ & The critic networks\\
% \hline
% $\Theta_{1_i}^{-}$, $\Theta_{2_i}^{-}$ & The target networks\\
% \hline
% $L(\cdot)$ & The loss function of value\\
% \hline
% $\mathcal{D}$ & Replay buffer\\
% \hline
% $V_{\Theta_{1_i}^{-}}(\cdot)$, $V_{\Theta_{2_i}^{-}}(\cdot)$ & State value functions\\
% \hline
% $\tau_i$ & Entropy regularity coefficient\\
% \hline
% \end{tabular}}
% \end{table}

The remainder of this paper is structured as follows. In Section \ref{se:2}, we provide a comprehensive description of the system model. Section \ref{se:3} presents the principles and details of the proposed algorithm, followed by the experimental results in Section \ref{se:4}. Finally, Section \ref{se:5} concludes the paper and outlines potential directions for future research.  Additionally, explanations of mainly used symbols are listed in Table \ref{tab:1}.

\begin{table}[!t]
\centering
\caption{Main Symbols and Explanations.\label{tab:1}}
\setlength{\tabcolsep}{2.0mm}{
\begin{tabular}{c|c}
\hline
{\bf Symbols} & {\bf Definition}\\
\hline\hline
$\mathbf{E}$ & Electric field strength \\
\hline
$Q_{enc}$ & The total charge\\
\hline
$\rho$ & The charge density\\
\hline
$\epsilon_0$ & Vacuum permittivity \\
\hline
$\delta$ & The dirac delta function\\
\hline
$V$ & The electric potential\\
\hline
$q$ & The point charge \\
\hline
$r$ & The distance from the point charge \\
\hline
$E_x$, $E_y$, $E_z$ & Electric field components \\
\hline
$(x_s,y_s,z_s)$ & The electric field source position\\
\hline
$Loss$ & The loss function\\
\hline
$\alpha$, $\beta$, $\gamma$ & The weighting coefficients\\
\hline
$b_l$ & The bias term \\
\hline
$h_k$ & The hidden layer node\\
\hline
$v_k$ & The visible layer node\\
\hline
$\sigma_{kl}$ & The connection weight matrix\\
\hline
$\mathcal{F}_a(\cdot)$ & The activation function\\
\hline
$\mathcal{z}_l$ & The hidden layer output\\
\hline
$\mathcal{D}$ & The training set\\
\hline
$\mathcal{G}$ & The gradient\\
\hline
$\mathcal{C}$ & The number of samples\\
\hline
$\mathcal{B}_1$, $\mathcal{B}_2$ & The exponential decay rates\\
\hline
$\mathcal{A}$ & The first moment variables\\
\hline
$\varsigma$ & The second moment variables\\
\hline
$H_k$ & The Hessian matrices from $k^{th}$ iterations\\
\hline
$I$ & The identity matrix\\
\hline
$s_k$ & The change in network's parameter\\
\hline
$g_k$ & The change in gradients\\
\hline
$\theta$ & The neural network's parameter\\
\hline
$u(E_X,E_y,E_z,\theta)$ & The neural network\\
\hline
\end{tabular}}
\end{table}

\begin{figure}[!t]
\centering
\includegraphics[width=0.948\linewidth]{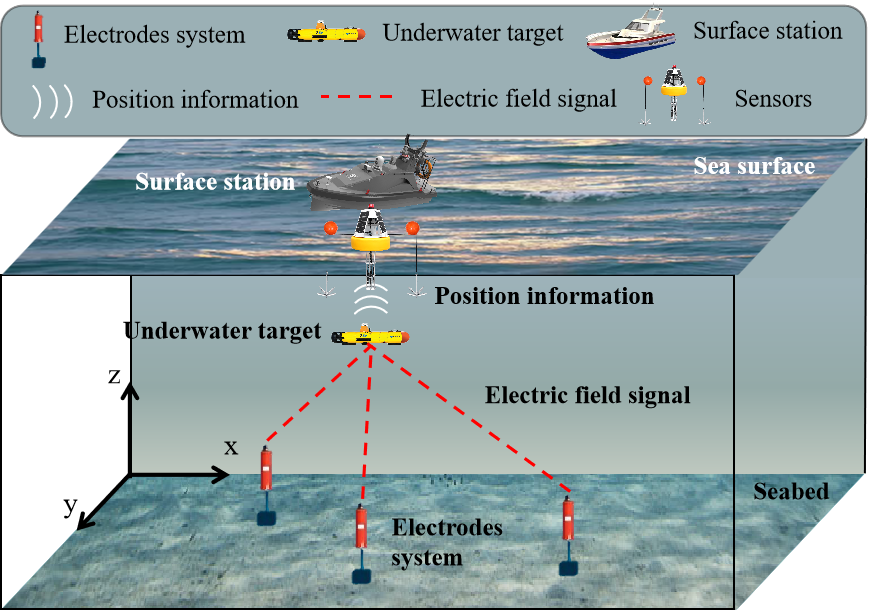}
\caption{Illustration of the underwater target localization scenario.}
\label{fig:1}
\end{figure}

\section{System Model}\label{se:2}

In this section, we provide a detailed description of the underwater localization task and the principles of electric field inversion. The subsequent algorithm development will be based on these system models.

\subsection{The Task Scenario of Underwater Localization}\label{se:2.1}

We consider an underwater target localization scenario, as illustrated in Fig. \ref{fig:1}, where a target, denoted as $T$, is arbitrarily located within a 3D space defined by $x$, $y$, and $z$ coordinates. Its position can be inferred through the inversion of electric field signals received by an electric field sensor network deployed on the target. Specifically, $N$ electrode systems are deployed at various underwater locations, generating an electric field by applying a voltage between the electrodes. The target (such as an Autonomous Underwater Vehicle, or AUV) obtains the $x$, $y$, and $z$ components of the electric field through onboard electric field sensors and inputs them into the EFILN localization system. EFILN performs inversion calculations on the input electric field signals and outputs the position coordinates of the target. These coordinates are then transmitted to a ground station via communication equipment, enabling the localization of the target.

\begin{figure*}
\centering
\includegraphics[width=0.948\linewidth]{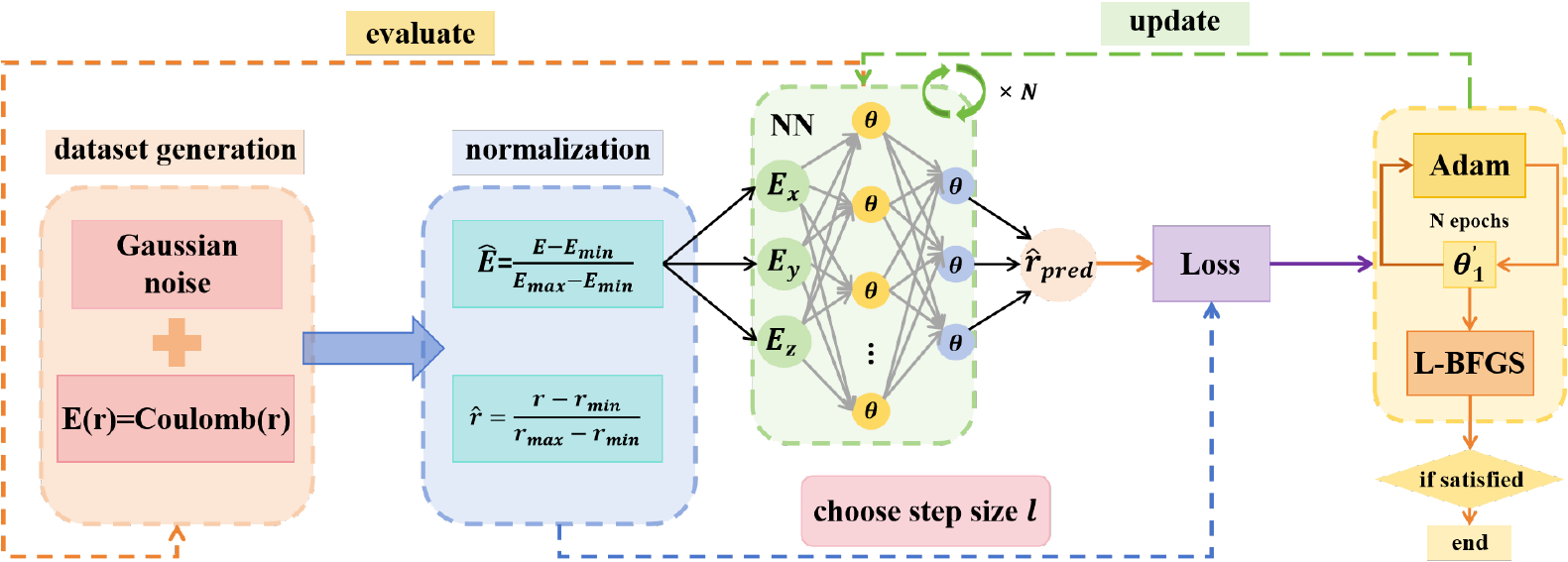}
\caption{The overall architecture of our proposed EFILN algorithm.}
\label{fig:2}
\end{figure*}

\subsection{Principles of Electric Field Localization}\label{se:2.2}

In electrostatics, Gauss's law from Maxwell's equations provides the relationship between the total charge enclosed within a closed surface and the integral of the electric flux over that surface:
\begin{equation}\oint_{\partial V}\mathbf{E}\cdot d\mathbf{A}=\frac{Q_{\mathrm{enc}}}{\epsilon_0},\end{equation}
where $\mathbf{E}$ is the electric field strength, and $d\mathbf{A}$ is the differential area element vector, whose direction is aligned with the normal to the surface. $\partial V$ is the closed surface in space $V$, $Q_{\mathrm{enc}}$ is the total charge enclosed by the surface, and $\epsilon_0$ is the vacuum permittivity.

According to the Gauss-Ostrogradsky theorem, the electric flux through a closed surface can be related to the volume integral of the divergence of the electric field within that volume:
\begin{equation}\oint_{\partial V}\mathbf{E}\cdot d\mathbf{A}=\int_V(\nabla\cdot\mathbf{E})dV.\end{equation}

Meanwhile, the total charge $Q_{\mathrm{enc}}$ enclosed by a closed surface can be expressed as the volume integral of the charge density $\rho$:
\begin{equation}Q_{\mathrm{enc}}=\int_V\rho dV.\end{equation}

Substituting Equation (2) and Equation (3) into Equation (1) yields:
\begin{equation}\int_V(\nabla\cdot\mathbf{E})dV=\frac1{\epsilon_0}\int_V\rho dV,\end{equation}
thus
\begin{equation}\frac{\partial E_x(x,y,z)}{\partial x}+\frac{\partial E_y(x,y,z)}{\partial y}+\frac{\partial E_z(x,y,z)}{\partial z}=\frac{\rho}{\epsilon_0}.\end{equation}
 
In this paper, since the electric field generated by the underwater electrode system can be approximated as a point charge, Equation (5) can be simplified. For a point charge $q$, its charge density $\rho$ can be expressed using the Dirac $\delta$ function:
\begin{equation}\rho=q\delta,\end{equation}
substituting Equation (6) into Equation (5) gives:
\begin{equation}\nabla^2V=-\frac{q\delta}{\epsilon_0},\end{equation}
where electric potential $V$ satisfies the equation $\mathbf{E} = -\nabla V$. Since the electric potential $V$ depends solely on the distance $r$ from the point charge, it can be assumed that the potential exhibits spherical symmetry. In this case, Equation (7) can be simplified to:
\begin{equation}\frac1{r^2}\frac d{dr}\left(r^2\frac{dV}{dr}\right)=-\frac{q\delta}{\epsilon_0}.\end{equation}

Based on the integral result from Gauss's law, the final solution for the electric potential distribution of a point charge is obtained as:
\begin{equation}V=\frac q{4\pi\epsilon_0r}.\end{equation}

Therefore, the corresponding relationship between the electric field distribution in space and the spatial coordinates is:
\begin{equation}\mathbf{E}(r)=-\frac{dV}{dr}\hat{\mathbf{r}}=-\frac d{dr}\left(\frac q{4\pi\epsilon_0r}\right)\hat{\mathbf{r}}=\frac q{4\pi\epsilon_0r^2}\hat{\mathbf{r}},\end{equation}
where $r=\sqrt{(x-x_s)^2+(y-y_s)^2+(z-z_s)^2}$, $\hat{\mathbf{r}}=(\frac{x-x_s}{r},\frac{y-y_s}{r},\frac{z-z_s}{r})$ is the unit vector from the spatial position $(x,y,z)$ to the electric field source position $(x_s,y_s,z_s)$. Equation (10) is the Coulomb law, which determines the electric field at a specific point based on spatial coordinates. However, inverting this equation using traditional numerical methods to derive spatial coordinates from electric field values is challenging. In this paper, we propose a neural network that takes electric field signals as input and outputs the corresponding spatial coordinates. The network's loss is calculated using Coulomb's law and iteratively optimized for high-precision localization. Specific details will be discussed in the next section.

\section{Algorithm Design}\label{se:3}

In this section, we first introduce the design of the algorithm's loss function, including how it is integrated with Coulomb law. Based on this, the principles and implementation details of the EFILN algorithm are presented, along with the corresponding pseudocode.

\subsection{Loss Function Design}\label{se:3.1}
 
Assume that $N$ electrode systems are deployed in the studied underwater area, with their coordinates given by $(x_s^i,y_s^i.z_s^i),i=1,\dots,N$. According to Equation (10), the electric field components in the three directions at the point $(x,y,z)$ can be calculated as follows:
\begin{equation}
    E_x=\sum_{i=1}^N(\frac {Q_i(x-x_s^i)}{4\pi\epsilon_0((x-x_s^i)^2+(y-y_s^i)^2+(z-z_s^i)^2)^{\frac{3}{2}}}),
\end{equation}

\begin{equation}
    E_y=\sum_{i=1}^N(\frac {Q_i(y-y_s^i)}{4\pi\epsilon_0((x-x_s^i)^2+(y-y_s^i)^2+(z-z_s^i)^2)^{\frac{3}{2}}}),
\end{equation}

\begin{equation}
    E_z=\sum_{i=1}^N(\frac {Q_i(z-z_s^i)}{4\pi\epsilon_0((x-x_s^i)^2+(y-y_s^i)^2+(z-z_s^i)^2)^{\frac{3}{2}}}),
\end{equation}
where $Q_i(i=1,\dots,N)$ represents the charge of the $i^{th}$ electrode system. 

Let $u(E_x,E_y,E_z,\theta)$ represents the neural network (NN) in the EFILN algorithm system. This network can take electric field signals as input and output spatial coordinates, thereby achieving electric field inversion localization. Assuming the dataset size is $M$, the loss in each of the three directions can be defined as follows:
\begin{equation}
    Loss_x=\frac{1}{M}\sum_{j=1}^M(u_x(E_x^j,E_y^j,E_z^j,\theta)-x_j)^2,
\end{equation}
\begin{equation}
    Loss_y=\frac{1}{M}\sum_{j=1}^M(u_y(E_x^j,E_y^j,E_z^j,\theta)-y_j)^2,
\end{equation}
\begin{equation}
    Loss_z=\frac{1}{M}\sum_{j=1}^M(u_z(E_x^j,E_y^j,E_z^j,\theta)-z_j)^2,
\end{equation}
where $u_x$, $u_y$, and $u_z$ represent the $x$, $y$ and $z$ coordinates of the position vector output by the NN, respectively. $x_j$, $y_j$, $z_j$, $E_x^j$, $E_y^j$ and $E_z^j$ correspond to the spatial coordinates and electric field information of the $j^{th}$ point in the dataset, respectively. $\theta$ is the neural network's parameter.

Based on the above analysis, our loss function is designed as follows:
\begin{equation}
    Loss=\alpha Loss_x+\beta Loss_y+\gamma Loss_z,
\end{equation}
where $\alpha$, $\beta$ and $\gamma$ are the weighting coefficients for the three components of the loss function.

\subsection{The EFILN Algorithm}\label{se:2.3}

The framework of the EFILN algorithm is illustrated in Fig. \ref{fig:2}. The challenge inherent in the localization algorithm resides in establishing a precise correspondence between the electric field signals and the spatial coordinates of the underwater target, culminating in the accurate derivation of said coordinates from the physical quantities. In this study, we utilize the NN to solve this challenge. Gaussian noise of a certain intensity is added to the dataset generated by the forward simulation using Coulomb law to better simulate real underwater environments. The noisy dataset is then normalized as follows: 
\begin{equation}
    \hat{\mathbf{X}}=(\frac{x-x_{min}}{x_{max}-x_{min}},\frac{y-y_{min}}{y_{max}-y_{min}},\frac{z-z_{min}}{z_{max}-z_{min}}),
\end{equation}

\begin{equation}
    \hat{\mathbf{E}}=(\frac{E_x-E_{xmin}}{E_{xmax}-E_{xmin}},\frac{E_y-E_{ymin}}{E_{ymax}-E_{ymin}},\frac{E_z-E_{zmin}}{E_{zmax}-E_{zmin}}),
\end{equation}
where $x_{min}$, $y_{min}$, $z_{min}$, $x_{max}$, $y_{max}$ and $z_{max}$ represent the minimum and maximum values of $x$, $y$ and $z$, respectively. The same applies to the parameters related to the electric field.

Then, the normalized input data undergoes a linear transformation through multiplication with weights and addition of a bias term $b_l$, yielding the layer's output. This process is mathematically represented as follows:
\begin{equation}\label{eq:26}
v_k=\sum_k\sigma_{kl}h_k+b_l,\ k=1,\cdots,n; l=1,\cdots,m,
\end{equation}
where $h_k$ denotes the hidden layer node ($0 <k\leq n$), $v_k$ is the visible layer node ($0 < k\leq m$), and $\sigma_{kl}$ represents the connection weight matrix between the visible layer and the hidden layer. The hidden layer output can be calculated as follows:
\begin{equation}\label{eq:27}
\mathcal{z}_l=\mathcal{F}_a(\sum_k\sigma_{kl}h_k+b_l),
\end{equation}
where $\mathcal{F}_a(\cdot)$ is the activation function. In this paper, the activation function is chosen to be tanh.

Given its fast convergence properties, we first utilize the Adam, an optimizer combining Momentum and RMSProp benefits. By sampling a minibatch of $\mathcal{J}$ examples from the training set $\{\mathcal{D}^{(1)},\ldots,\mathcal{D}^{(\mathcal{J})}\}$ with corresponding targets $\mathcal{I}^{(\mathcal{V})}$, the gradient can be computed as
\begin{equation}\label{eq:33}
\mathcal{G}\leftarrow\frac1C\nabla_{\theta}\sum_{\mathcal{V}=1}^{\mathcal{J}}Loss(u(\mathcal{D}^{(\mathcal{V})},\theta),\mathcal{I}^{(\mathcal{V})}),
\end{equation}where $\mathcal{G}$ is the gradient and $C$ is the number of samples in a batch.

Then, we update biased first and second moment estimate
\begin{equation}\label{eq:34}
\mathcal{A}\leftarrow \mathcal{B}_1\mathcal{A}+(1-\mathcal{B}_1\mathcal{G}),
\end{equation}
\begin{equation}\label{eq:35}
\varsigma\leftarrow \mathcal{B}_2\varsigma+(1-\mathcal{B}_2\mathcal{G})\mathcal{G},
\end{equation}
where $\mathcal{B}_{1}$ and $\mathcal{B}_{2}$ are the exponential decay rates for moment estimates, and $\mathcal{A}$ and $\varsigma$ are the first and second moment variables, respectively. Hence, it is possible to rectify bias in both the first and second moments through the computation of the update. The process of achieving the gradient update entails the repetition of the aforementioned cycle.

After initial optimization with the Adam optimizer, we switched to the L-BFGS optimizer for fine-tuning. L-BFGS is a quasi-Newton method that updates parameters based on an approximation of the second-order derivative (Hessian matrix). It excels at making fine adjustments in the later training stages, helping the model reach local optima, and is also memory efficient. The update formula is as follows:
\begin{equation}H_{k+1}=\left(I-\frac{s_kg_k^T}{g_k^Ts_k}\right)H_k\left(I-\frac{g_ks_k^T}{g_k^Ts_k}\right)+\frac{s_ks_k^T}{g_k^Ts_k},\end{equation}
where $H_k$ and $H_{k+1}$ represent the Hessian matrices from the $k^{th}$ and $(k+1)^{th}$ iterations, respectively. $I$ is the identity matrix, $s_k=\theta_{k+1}-\theta_k$ denotes the change in network's parameter, and $g_k=\nabla u(\theta_{k+1})-\nabla u(\theta_k)$ represents the change in gradients. L-BFGS limits memory usage by storing only the updates from the most recent iterations to construct an approximation of the Hessian matrix. Our optimization strategy, which first uses Adam followed by L-BFGS, combines the strengths of both: Adam's efficient exploration and L-BFGS's precise optimization. The pseudocode for EFILN is presented in Algorithm \ref{alg:1}.

\vspace{-0.2cm}
\begin{algorithm}[!t]
\label{alg:1}
\caption{The EFILN Algorithm}
Initialize the neural network's parameter $\theta$.

Sampling points at specified intervals $x_{step},y_{step},z_{step}$ and applying Coulomb law to generate the position-electric field dataset. Then normalized this dataset.

\For{each epoch $n$}{

Input the electric field information from the dataset into the neural network and output the predicted position coordinates $(x_{pred},y_{pred},z_{pred})$.

Calculate the loss by Coulomb law:
\begin{equation}
    Loss=\alpha Loss_x+\beta Loss_y+\gamma Loss_z.
\end{equation}

Updated parameters by Adam algorithm.
}
Update parameters by L-BFGS until the gradient tolerance or change tolerance falls below the specified threshold.

Evaluate the effects of EFILN.

\end{algorithm}
\vspace{0.2cm}

\section{Experiments and Results}\label{se:4}

In this section, we first present the experiment settings, including both environmental and algorithmic parameters. Subsequently, simulations under various conditions are conducted to validate the superior performance of EFILN.

\subsection{Experiment Settings}

The experimental code in this study was executed on a personal computer equipped with 13th Gen Intel® Core™ i7-13650HX processor and NVIDIA GeForce RTX 4060 Laptop GPU. The parameters can be divided into two parts: environment parameters and algorithm parameters, which are listed in Table \ref{tab:2} for summary.

% \begin{figure}[!t]
% \centering
% \includegraphics[width=0.948\linewidth]{mxsyt.pdf}
% \caption{Three-layeThe experimental code in this study was executed on a personal computer equipped with 13th Gen Intel® Core™ i7-13650HX processor and NVIDIA GeForce RTX 4060 Laptop GPU. The experimental parameters include simulation environment parameters, statistical signal processing model parameters, and hyperparameters of the RL algorithm (using the soft actor-critic algorithm, SAC, as an example in this study), which are listed in Table 1, Table 2, and Table 3, respectively.r model of the atmosphere, ocean, and seabed sediments.}
% \label{fig:4}
% \end{figure}

\begin{table}[!t]
\caption{Parameters of the environment and fusion framework.\label{tab:2}}
\centering
\setlength{\tabcolsep}{0.5mm}{
\begin{tabular}{m{4.7cm}<{\centering}c}
\hline
{\bf Parameters} & {\bf Values}\\
\hline
Permittivity of vacuum $\epsilon_0$ & 8.854$\times$$10^{-12}$ $F/m$ \\
$x,y,z$-direction underwater range & [10$m$, 110$m$]\\
Step size & 0.5 $m$\\
The number of electrode systems & 2 \\
The amount of charge on the electrode & $\pm$1 $C$\\
The positions of the electrode systems & (0,0,0), 
 (0,0,100)\\
The number of neurons in hidden layers & 16 \\
The number of hidden layers & 8 \\
The learning rate of Adam & 0.0001\\
The learning rate of L-BFGS & 10\\
The history size of L-BFGS & 50\\
The gradient tolerance of the L-BFGS & 1$\times$$10^{-12}$ \\
The maximum number of epochs & 50000\\
\hline
\end{tabular}}
\end{table}

\begin{figure*}[!t]
	
	\begin{minipage}{0.32\linewidth}
		\vspace{3pt}
		\centerline{\includegraphics[width=\textwidth]{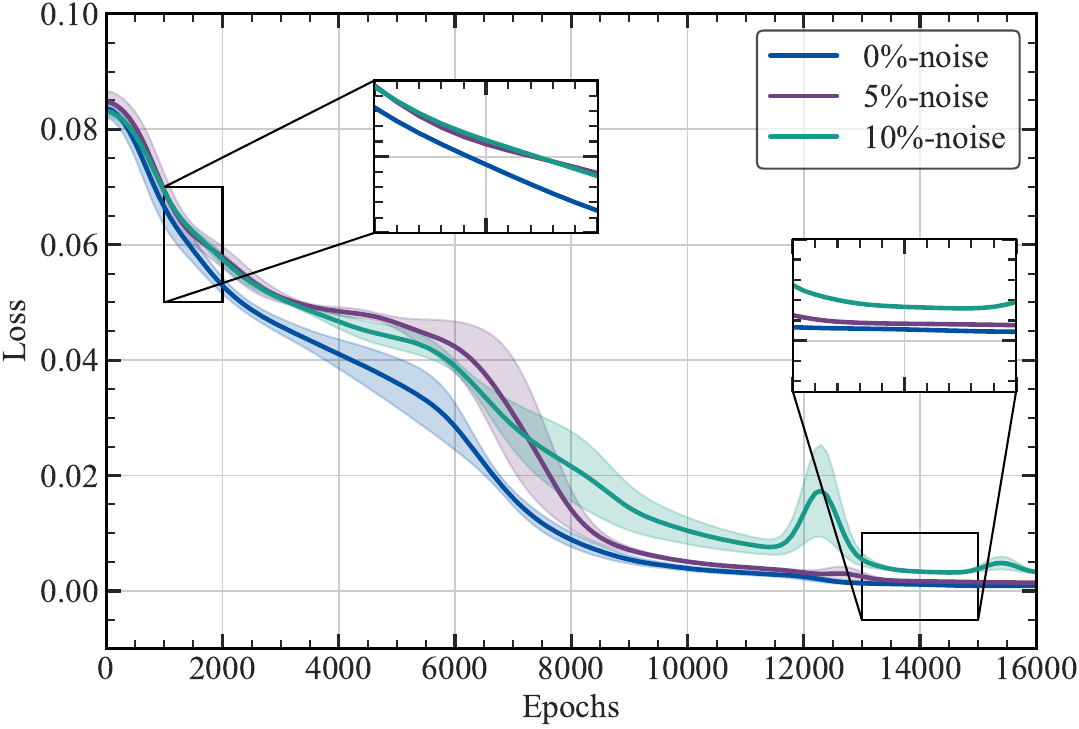}}
          % 加入对这列的图片说明
		\centerline{(a) x-direction}
	\end{minipage}
	\begin{minipage}{0.32\linewidth}
		\vspace{3pt}
		\centerline{\includegraphics[width=\textwidth]{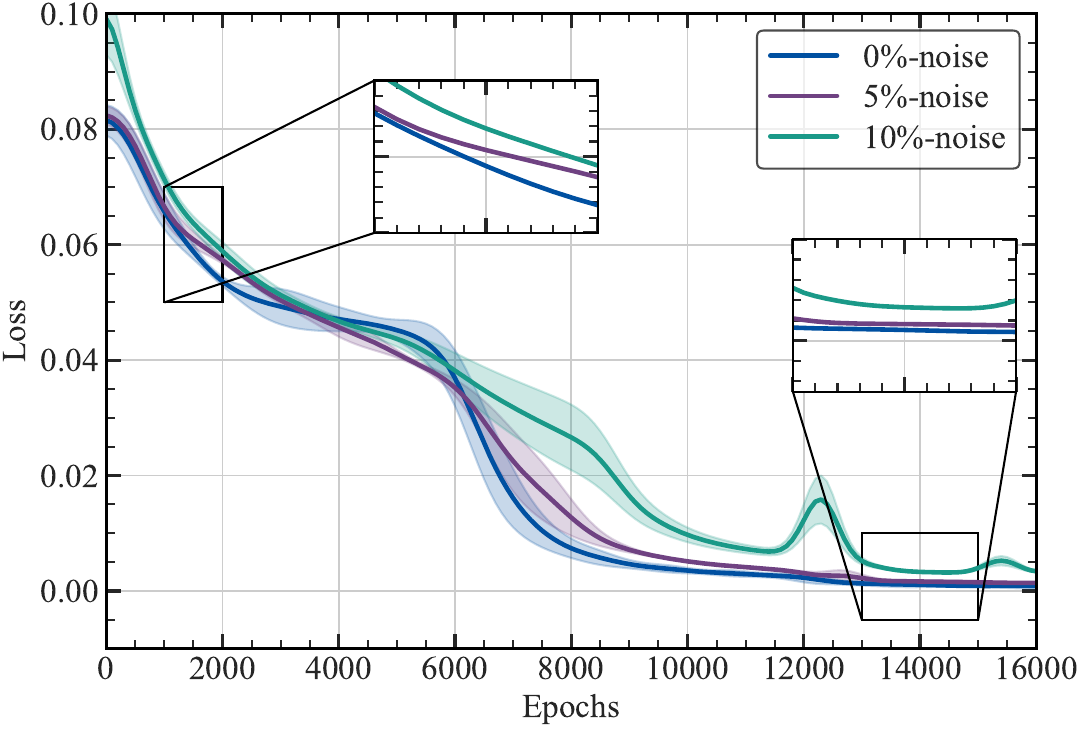}}
	 
		\centerline{(b) y-direction}
	\end{minipage}
	\begin{minipage}{0.32\linewidth}
		\vspace{3pt}
		\centerline{\includegraphics[width=\textwidth]{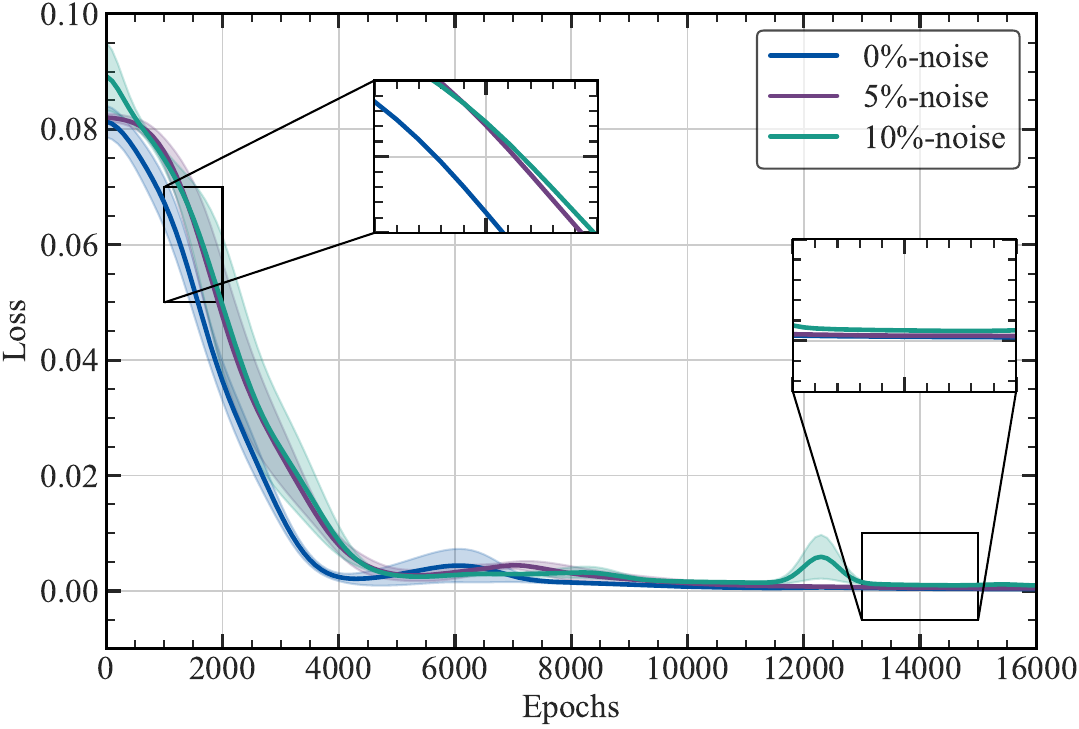}}
	 
		\centerline{(c) z-direction}
	\end{minipage}
 
	\caption{Comparison of three-directional training loss under different noise levels: (a) x-direction's training loss (b) y-direction's training loss (c) z-direction's training loss.}
	\label{fig:3}
\end{figure*}

\begin{figure*}[!t]
	
	\begin{minipage}{0.32\linewidth}
		\vspace{3pt}
		\centerline{\includegraphics[width=\textwidth]{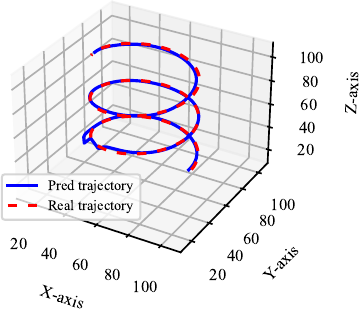}}
          % 加入对这列的图片说明
		\centerline{(a) spiral trajectory}
	\end{minipage}
	\begin{minipage}{0.32\linewidth}
		\vspace{3pt}
		\centerline{\includegraphics[width=\textwidth]{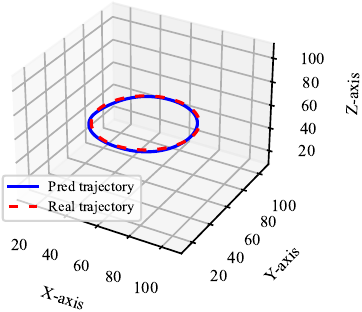}}
	 
		\centerline{(b) circular trajectory}
	\end{minipage}
	\begin{minipage}{0.32\linewidth}
		\vspace{3pt}
		\centerline{\includegraphics[width=\textwidth]{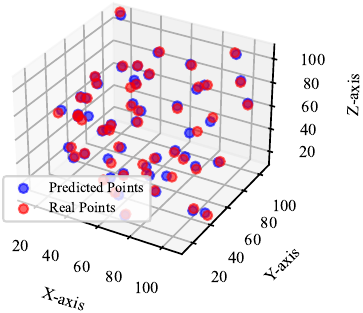}}
	 
		\centerline{(c) randomly selected points}
	\end{minipage}
 
	\caption{Comparison of real and predicted trajectories under three different simulation paths: (a) spiral trajectory (b) circular trajectory (c) randomly selected points.}
	\label{fig:4}
\end{figure*}

\subsection{Experiment Results}

\begin{table}[!t]
    \centering
    \caption{Comparison of three-directional localization accuracy under different noise intensities.}
    \label{tab:3}
    \begin{tabular}{lcccc}
    \toprule
      & 0\% & 5\% & 10\%  \\
    \midrule
    $x$ & 0.814$m$ & 1.013$m$ & 1.274$m$ \\
    $y$ & 0.908$m$ & 1.266$m$ & 0.970$m$ \\
    $z$ & 0.863$m$ & 0.730$m$ & 0.803$m$ \\
    \bottomrule
    \end{tabular}
\end{table}

\begin{table}[!t]
    \centering
    \caption{Localization accuracy comparison across varying sampling intervals.}
    \label{tab:4}
    \begin{tabular}{lcccc}
    \toprule
      & 0.5$m$ & 0.75$m$ & 1.0$m$  \\
    \midrule
    $x$ & 0.814$m$ & 1.102$m$ & 1.328$m$ \\
    $y$ & 0.908$m$ & 1.345$m$ & 1.651$m$ \\
    $z$ & 0.863$m$ & 1.182$m$ & 0.948$m$ \\
    \bottomrule
    \end{tabular}
\end{table}

\begin{table}[!t]
    \centering
    \caption{Comparison of different localization algorithms.}
    \label{tab:5}
    \begin{tabular}{lcccc}
    \toprule
      & EFILN & RL- glmu & CSPLLS  \\
    \midrule
    location error & 0.9\% & 2.0\% & 10.0\% \\
    \bottomrule
    \end{tabular}
\end{table}

To better simulate the complexity of underwater environments and verify robustness, Gaussian noise with intensities of 5\% and 10\% was added to the dataset. The comparison of the three-directional loss under different noise conditions and the localization accuracy are shown in Fig. \ref{fig:3} and Table \ref{tab:3}, respectively. As noise increases, the dataset becomes more disrupted, leading to higher training loss. However, even in the presence of noise, the loss of EFILN still converges to a very low value, and it achieves high localization accuracy, confirming the robustness of EFILN.

Subsequently, small-sample tests were conducted to evaluate the SSL capability of EFILN. Starting with a sampling interval of 0.5$m$, the step size was gradually increased by 0.25$m$ during the training process, reaching a final value of 1.0$m$. The experimental results are shown in Table \ref{tab:4}. It can be observed that as the sampling interval increases, the positioning accuracy of EFILN decreases slightly. This is because the increase in step size leads to a significant reduction in the dataset size, thus weakening the neural network’s fitting ability. However, even with a limited number of samples, EFILN still achieves high positioning accuracy, demonstrating the strong SSL capability of the algorithm, which meets the positioning accuracy requirements to some extent even with insufficient samples.

To visualize EFILN's high-precision localization capability, its performance was validated using spiral, circular and randomly selected spatial trajectories. To ensure rigorous validation, the selected points for the three types of trajectories do not overlap with any points in the dataset. The results are shown in Fig. \ref{fig:4}. The predicted trajectories closely overlap with the real trajectories in all three cases, demonstrating the practical utility of EFILN.

Finally, under the conditions of the simulation experiments in this paper, the localization error of EFILN was compared with two other localization algorithms, RL-glmu\cite{10400257} and CSPLLS\cite{8795492}. The results, shown in Table \ref{tab:5}, indicate that EFILN's localization error is less than 50\% and 10\% of the other two methods, respectively, further demonstrating the superior performance of EFILN.

\section{Conclusion}\label{se:5}

In this paper, we developed a deep neural network, EFILN, based on electric field signals for high-precision underwater localization. This network leverages the relationship between electric field signals generated by underwater electrode systems and spatial positions to invert the electric field data and accurately determine position coordinates. By employing a combination of the Adam and L-BFGS optimizers, the network's training performance is further enhanced. Extensive simulation experiments demonstrate that the network achieves high localization accuracy, exhibits strong robustness, and is well-suited for small-sample learning. Considering the complexity of underwater environments, future work will focus on validating real-world underwater field data.

\bibliographystyle{IEEEtran}
%\bibliography{WCNC.bib}

\begin{thebibliography}{10}
\providecommand{\url}[1]{#1}
\csname url@samestyle\endcsname
\providecommand{\newblock}{\relax}
\providecommand{\bibinfo}[2]{#2}
\providecommand{\BIBentrySTDinterwordspacing}{\spaceskip=0pt\relax}
\providecommand{\BIBentryALTinterwordstretchfactor}{4}
\providecommand{\BIBentryALTinterwordspacing}{\spaceskip=\fontdimen2\font plus
\BIBentryALTinterwordstretchfactor\fontdimen3\font minus \fontdimen4\font\relax}
\providecommand{\BIBforeignlanguage}[2]{{%
\expandafter\ifx\csname l@#1\endcsname\relax
\typeout{** WARNING: IEEEtran.bst: No hyphenation pattern has been}%
\typeout{** loaded for the language `#1'. Using the pattern for}%
\typeout{** the default language instead.}%
\else
\language=\csname l@#1\endcsname
\fi
#2}}
\providecommand{\BIBdecl}{\relax}
\BIBdecl

\bibitem{khalil2020toward}
R.~A. Khalil, N.~Saeed, M.~I. Babar, and T.~Jan, ``Toward the internet of underwater things: Recent developments and future challenges,'' \emph{IEEE Consumer Electronics Magazine}, vol.~10, no.~6, pp. 32--37, 2020.

\bibitem{pan2021iout}
X.~Pan, Y.~Shen, and J.~Zhang, ``Iout based underwater target localization in the presence of time synchronization attacks,'' \emph{IEEE Transactions on Wireless Communications}, vol.~20, no.~6, pp. 3958--3973, 2021.

\bibitem{su2022algorithm}
R.~Su, Z.~Gong, C.~Li, and X.~Shen, ``Algorithm design and performance analysis of target localization using mobile underwater acoustic array networks,'' \emph{IEEE Transactions on Vehicular Technology}, vol.~72, no.~2, pp. 2395--2406, 2022.

\bibitem{lee2018real}
H.~Lee, H.-K. Jung, S.-H. Cho, Y.~Kim, H.~Rim, and S.~K. Lee, ``Real-time localization for underwater moving object using precalculated dc electric field template,'' \emph{IEEE Transactions on Geoscience and Remote Sensing}, vol.~56, no.~10, pp. 5813--5823, 2018.

\bibitem{7382734}
J.~Dou, S.~Chao-long, W.~Xiang-jun, and W.~Zi-xia, ``Research on electric field localization algorithm based on kalman filter,'' in \emph{2015 Chinese Automation Congress (CAC)}, 2015, pp. 1485--1488.

\bibitem{Yu2019SHIPTT}
\BIBentryALTinterwordspacing
P.~Yu, J.~Cheng, and J.~Zhang, ``Ship target tracking using underwater electric field,'' \emph{Progress In Electromagnetics Research M}, 2019. [Online]. Available: \url{https://api.semanticscholar.org/CorpusID:214465529}
\BIBentrySTDinterwordspacing

\bibitem{liu2021two}
W.~Liu, Z.~Xi, H.~Wang, and R.~Zhang, ``Two-dimensional deep learning inversion of magnetotelluric sounding data,'' \emph{Journal of Geophysics and Engineering}, vol.~18, no.~5, pp. 627--641, 2021.

\bibitem{shi1998one}
X.~Shi and J.~Wang, ``One dimensional magnetotelluric sounding inversion using simulated annealing,'' \emph{Earth Science-Journal of China University of Geosciences (in Chinese)}, vol.~23, no.~5, pp. 542--546, 1998.

\bibitem{sharma2012vfsares}
S.~P. Sharma, ``Vfsares--a very fast simulated annealing fortran program for interpretation of 1-d dc resistivity sounding data from various electrode arrays,'' \emph{Computers \& Geosciences}, vol.~42, pp. 177--188, 2012.

\bibitem{shaw2007particle}
R.~Shaw and S.~Srivastava, ``Particle swarm optimization: A new tool to invert geophysical data,'' \emph{Geophysics}, vol.~72, no.~2, pp. F75--F83, 2007.

\bibitem{shi2009damped}
X.-M. SHI, M.~XIAO, J.-K. FAN, G.-S. YANG, and X.-H. ZHANG, ``The damped pso algorithm and its application for magnetotelluric sounding data inversion,'' \emph{Chinese Journal of Geophysics}, vol.~52, no.~4, pp. 1114--1120, 2009.

\bibitem{schwarzbach2005two}
C.~Schwarzbach, R.-U. B{\"o}rner, and K.~Spitzer, ``Two-dimensional inversion of direct current resistivity data using a parallel, multi-objective genetic algorithm,'' \emph{Geophysical Journal International}, vol. 162, no.~3, pp. 685--695, 2005.

\bibitem{10650551}
J.~Xu, Y.~Ding, Z.~Zhang, G.~Xie, Z.~Wang, Y.~Zeng, and G.~Li, ``Multi-auv assisted seamless underwater target tracking relying on deep learning and reinforcement learning,'' in \emph{2024 International Joint Conference on Neural Networks (IJCNN)}, 2024, pp. 1--9.

\bibitem{inproceedings}
Y.~Ding, J.~Xu, G.~Xie, G.~Li, J.~Wang, and Y.~Ren, ``Advanced framework for underwater node repair via multi-auv based on multi-agent offline reinforcement learning,'' 09 2024.

\bibitem{puzyrev2019deep}
V.~Puzyrev, ``Deep learning electromagnetic inversion with convolutional neural networks,'' \emph{Geophysical Journal International}, vol. 218, no.~2, pp. 817--832, 2019.

\bibitem{di2020new}
Q.~Di, G.~Xue, C.~Yin, and X.~Li, ``New methods of controlled-source electromagnetic detection in china,'' \emph{Science China Earth Sciences}, vol.~63, pp. 1268--1277, 2020.

\bibitem{liu2020deep}
B.~Liu, Q.~Guo, S.~Li, B.~Liu, Y.~Ren, Y.~Pang, X.~Guo, L.~Liu, and P.~Jiang, ``Deep learning inversion of electrical resistivity data,'' \emph{IEEE Transactions on Geoscience and Remote Sensing}, vol.~58, no.~8, pp. 5715--5728, 2020.

\bibitem{10400257}
Y.~Yue, W.~Su, and Y.~Zhao, ``Underwater acoustic aided grid localization of multi-uuvs with reinforcement learning,'' in \emph{2023 IEEE International Conference on Signal Processing, Communications and Computing (ICSPCC)}, 2023, pp. 1--6.

\bibitem{8795492}
G.~Qiao, C.~Zhao, F.~Zhou, and N.~Ahmed, ``Distributed localization based on signal propagation loss for underwater sensor networks,'' \emph{IEEE Access}, vol.~7, pp. 112\,985--112\,995, 2019.

\end{thebibliography}

%\newpage

%needed around the contents of the optional argument to biography to prevent
 %the LaTeX parser from getting confused when it sees the complicated
 %your own custom macro containing the $\backslash${\tt{includegraphics}} command to make things
 

%\bf{If you include a photo:}\vspace{-33pt}
%\begin{IEEEbiography}[{\includegraphics[width=1in,height=1.25in,clip,keepaspectratio]{fig1}}]{Michael Shell}
%Use $\backslash${\tt{begin\{IEEEbiography\}}} and then for the 1st argument use $\backslash${\tt{includegraphics}} to %declare and link the author photo.
%Use the author name as the 3rd argument followed by the biography text.
%\end{IEEEbiography}




%\vfill

\end{document}